\documentclass[prb,superscriptaddress,twocolumn]{revtex4}
\usepackage{graphicx}
\usepackage{amsmath}
\usepackage{amssymb}
\usepackage{color}

\begin{document}

%\normalem

\title{Coexistence and competition of nematic and gapped states in bilayer graphene}

\date{September 8, 2012}

%\preprint{UWO-TH-12/..}

\author{E. V. Gorbar}
%\email{gorbar@bitp.kiev.ua}
\affiliation{Department of Physics, Taras Shevchenko National Kiev University, Kiev, 03022, Ukraine}
\affiliation{Bogolyubov Institute for Theoretical Physics, Kiev, 03680, Ukraine}

\author{V. P. Gusynin}
%\email{vgusynin@bitp.kiev.ua}
\affiliation{Bogolyubov Institute for Theoretical Physics, 03680, Kiev, Ukraine}

\author{V. A. Miransky}
%\email{vmiransk@uwo.ca}
\affiliation{Department of Applied Mathematics, Western University,
London, Ontario, Canada,  N6A 5B7}

\author{I. A. Shovkovy}
%\email{igor.shovkovy@asu.edu}
\affiliation{Department of Applied Sciences and Mathematics, Arizona State
University, Mesa, Arizona 85212, USA}

\begin{abstract}
In bilayer graphene, the phase diagram in the plane of a strain-induced bare nematic term, ${\cal N}_{0}$,
and a top-bottom gates voltage imbalance, $U_{0}$, is obtained by solving the gap equation in the
random-phase approximation. At nonzero ${\cal N}_0$ and $U_0$, the phase diagram consists
of two hybrid spin-valley symmetry-broken phases with both nontrivial nematic and mass-type order
parameters. The corresponding phases are separated by a critical line of first- and second-order
phase transitions at small and large values of ${\cal N}_0$, respectively. The existence of a critical end
point, where the line of first-order phase transitions terminates, is predicted. For ${\cal N}_0=0$,
a pure gapped state with a broken spin-valley symmetry is the ground state of the system. As
${\cal N}_{0}$ increases, the nematic order parameter increases, and the gap weakens in the
hybrid state. For $U_{0}=0$, a quantum second-order phase transition from the hybrid state
into a pure gapless nematic state occurs when the strain reaches a critical value. A nonzero
$U_{0}$ suppresses the critical value of the strain. The relevance of these results to recent
experiments is briefly discussed.
\end{abstract}


%%% 76.40.+b
%%% 73.43-f
%%% 75.30.Ds

\maketitle

\section{Introduction}

At present, a significant amount of attention is being paid to the ground state of bilayer graphene at
the neutrality point. Various gapped states with broken spin-valley symmetry, such as quantum anomalous
Hall (QAH),\cite{QAH} quantum spin Hall (QSH),\cite{QSH} quantum valley Hall (QVH),\cite{QVH}
layer antiferromagnet (LAF),\cite{Zhang} as well as a gapless nematic state,\cite{Vafek,Lemonik}
were suggested as candidates for the ground state. Experiments\cite{Weitz,Freitag,Velasco1,Velasco2}
showed that bilayer graphene at the neutrality point is gapped in the absence of external fields.
However, the experiment performed by the Manchester group\cite{Mayorov} found a gapless
state, and there were strong indications that it was a nematic state.

A variety of suggested ground states is related to the way in which the approximate $SU(4)$ spin-valley
symmetry of the low energy effective model of bilayer graphene is broken in its ground state.
The application of external electric and magnetic fields adds more complexity to the problem
of the ground state of bilayer graphene. However, it opens the possibility to use them as useful
probes of the ground state of the system. Theoretical papers\cite{GGM,Levit,Toke} predicted
and experiments\cite{Weitz,Kim,Velasco1,Velasco2} showed the presence of a phase transition
between the QSH and the QVH states as an electric field increases. Since the QVH (layer
polarized) state is the ground state of the system for a sufficiently large electric field, this
rules out the QVH state as a candidate for the ground state in the absence of external fields.

Simple continuity arguments suggest that the LAF and QSH states are the most likely candidates
for the gapped ground state of bilayer graphene at the neutrality point in the absence of external
fields. Indeed, for a sufficiently large magnetic field, the QSH (spin polarized) state is the ground state
of bilayer graphene. On the other hand, the experiments\cite{Freitag,Velasco1,Velasco2} do not show
any phase transitions in bilayer graphene at the neutrality point as a magnetic field is switched off.
This excludes the QAH state because it cannot be smoothly connected to the spin polarized state
at large magnetic fields. The LAF state, on the other hand, can adiabatically evolve into the QSH
one\cite{Kharitonov} and, therefore, cannot be excluded.

In Refs.~\onlinecite{QVH} and \onlinecite{MacD}, it was shown that, unlike single-layer graphene with its linear dispersion 
relation, the quadratic dispersion relation in bilayer graphene leads to an instability of its normal phase 
for arbitrary weak Coulomb or short-range repulsive interactions. Renormalization group studies in 
the normal phase of bilayer graphene in Refs.~\onlinecite{Vafek} and \onlinecite{Lemonik} revealed 
a variety of instabilities in this phase, which can potentially lead to different (competing) ground 
states with condensates. It was also found that the instability with respect to the generation
of a nematic order parameter is strongest, which seems to suggest that the nematic state is
the ground state of bilayer graphene.  However, since the condensates essentially modify the gap equations,
in order to determine the phase diagram of the system, it is necessary to analyze the gap equations
for states with different condensates and then to compare their energy densities.
This is the main goal of the present paper.

Solving the gap equation in the random-phase approximation, we find  the phase
diagram in the plane of a strain-induced bare nematic term ${\cal N}_{0}$ and a top-bottom
gates voltage imbalance $U_{0}$. We show that the ground state of the system at
${\cal N}_{0}\ne0$ and $U_{0}\ne0$ is a hybrid state with nonzero nematic and mass-type
order parameters. The critical line separates the phase with one of the hybrid QAH, QSH, or
LAF states (which are degenerate in energy in the model at hand) from the hybrid QVH state.
The phase transition along a large part of this critical line is of second-order except for the region of small
${\cal N}_{0}$ where a sufficiently large value of $U_{0}$ drives the first-order phase transition
between the two different states with broken spin-valley symmetry. The predicted
existence of a critical end point in the phase diagram may be relevant to current experiments
in bilayer graphene.

The paper is organized as follows.
The analysis is performed in the framework of the two-band model of bilayer graphene with the Coulomb
interaction between quasiparticles described in Sec.~\ref{secII}. In Sec.~\ref{secIII} we derive the gap equations
in the random-phase approximation for the gapped and nematic order parameters. The main results
of the paper, including the numerical solutions to the gap equations, are presented in Sec.~\ref{secIV}.
Sec.~\ref{Summary} contains a brief summary of the results, as well as our discussions and conclusions.
In the Appendix the derivation of the expression for the energy density is given.

\section{Model}
\label{secII}

The free part of the effective low-energy Hamiltonian of bilayer graphene is as follows:\cite{McC}
\begin{equation}
H_0 = - \frac{1}{2m}\sum_{\xi, s}\int
d^2x\,\Psi_{\xi s}^+(x)\left(
\begin{array}{cc}
0 & (\pi^{\dagger})^2\\
\pi^2 & 0
\end{array}
\right)\Psi_{\xi s}(x),
\label{free-Hamiltonian}
\end{equation}
where $\pi=\hat{k}_{1}+i\hat{k}_{2}$ is the canonical momentum operator,
$m= \gamma_1/2v_{F}^2$, where the Fermi velocity is $v_F \simeq c/300$, and
$\gamma_1 \approx 0.4~\mbox{eV}$. The two-component spinor field $\Psi_{\xi s}$
carries the valley $(\xi=\pm$ for $K$ and $K^{\prime}$ valleys, respectively) and spin
($s = \pm$) indices. We will use the standard convention:\cite{McC} $\Psi_{+s}^T=
(\psi_{+A{_1}}, \psi_{+B{_2}})_s$, whereas, $\Psi_{-s}^T = (\psi_{-B{_2}},
\psi_{-A{_1}})_s$. Here, $A_1$ and $B_2$ correspond to those sublattices in 
layers 1 and 2, respectively, which, according to Bernal $(A_2-B_1)$ stacking,
are relevant for the low energy dynamics. Let us emphasize that the sublattice and
layer degrees of freedom are not independent in this low-energy model: The sublattices
$A_1$ and $B_2$ correspond to layers $1$ and $2$, respectively. The effective Hamiltonian
(\ref{free-Hamiltonian}) is valid up to energies $\Lambda=\gamma_1/4\approx 0.1~\mbox{eV}$
and we ignore small trigonal warping effects.\cite{McC}

The Coulomb interactions in bilayer graphene are given by\cite{QVH,GGM}
\begin{eqnarray}
&&\hspace{-5mm}H_{C}=\frac{1}{2}\int \hspace{-1.0mm}d^2xd^2x^{\prime}\Big\{
V(x-x^{\prime}) \left[ \rho_1(x) \rho_1(x^{\prime})+
\rho_2(x)\rho_2(x^{\prime})\right] \nonumber\\
&&
\hspace{25.0mm} +
2V_{12}(x-x^{\prime})\rho_1(x)\rho_2(x^{\prime}) \Big\},
\label{interaction}
\end{eqnarray}
where the interaction potentials $V (x)$ and $V_{12}(x)$ describe the intralayer and interlayer interactions,
respectively. Their Fourier transforms are ${V}(k)= 2\pi e^2/\kappa k$ and ${V}_{12}(k)= 2\pi e^2e^{-kd}/\kappa k$,
where $d\approx 0.35~\mbox{nm}$ is the distance between the layers, and $\kappa$ is a dielectric constant
(in the numerical analysis below, we set $\kappa=4$). The two-dimensional charge-density operators
$\rho_1(x)$ and $\rho_2(x)$ are
\begin{equation}
\rho_l (x)=\sum_{\xi,s=\pm} \Psi^+_{\xi s}(x) P_l (\hat{\xi}) \Psi_{\xi s}(x),\quad (l=1,2),
\label{density}
\end{equation}
where $P_1(\hat{\xi}) =(1+\hat{\xi}\tau^3)/2$ and $P_2(\hat{\xi})=(1-\hat{\xi}\tau^3)/2$ are projectors on states
in the layers 1 and 2, respectively. Here the matrix $\hat{\xi}$ acts as $\hat{\xi}\Psi_{\xi s}=\xi\Psi_{\xi s}$,
and $\tau^3$ is the diagonal Pauli matrix acting on the two components of the fields
$\Psi_{+ s}$ and $\Psi_{- s}$. Note that the presence of $\hat{\xi}$ in $P_1(\hat{\xi})$ and $P_2(\hat{\xi})$
is related to the opposite order of the $A_1$ and $B_2$ components in
$\Psi_{+ s}$ and $\Psi_{- s}$.

Whereas, the gaps in the QAH and QSH states in bilayer graphene are analogous to a Haldane
mass,\cite{QAH,QSH} the gaps in the QVH and LAF states are analogous to a Dirac
mass.\cite{Zhang,GGM} In bilayer graphene, the latter is realized as a voltage imbalance 
between the layers.\cite{GGM,McC} The corresponding order parameters (condensates) are 
as follows\cite{Zhang,GGM}: spin singlets $\langle\Psi^{\dagger}\tau^3\Psi\rangle$ (QAH)
and $\langle\Psi^{\dagger}\hat{\xi}\tau^3\Psi\rangle$ (QVH), and spin
triplets $\langle\Psi^{\dagger}\tau^3\sigma^3\Psi\rangle$ (QSH) and
$\langle\Psi^{\dagger}\hat{\xi}\tau^3\sigma^3\Psi\rangle$ (LAF), where $\sigma_3$ is
a spin Pauli matrix (the indices $\xi$ and $s$ in the field $\Psi_{\xi s}$ are omitted here).

Without a magnetic field, the QSH and QVH states are invariant under time reversal,\cite{QSH,QVH}
whereas, the QAH and LAF states break this symmetry\cite{QAH,Zhang} (the QAH state is associated
with the Chern number \cite{QAH} and the QSH state is a two-dimensional topological insulator associated
with the $Z_2$ topological invariant \cite{QSH}).
On the other hand, although the QAH state is invariant under the $SU(4)$, the QSH one breaks
the spin $SU(2)$ subgroup of $SU(4)$ down to $U(1)$.\cite{QSH,GGM} Both the QVH
and the LAF states break a $Z_2$ subgroup of the spin-valley $SU(4)$ describing the
valley transformation $\xi \to -\xi$ ($\Psi_{+s}\to \Psi_{-s}$),\cite{GGM} in
 addition to that, the LAF state breaks the spin $SU(2)$ down to $U(1)$.

A usual nematic order parameter breaks the $SO(2)$ rotational group down to a discrete
$Z_2$ subgroup. As we discuss below, however, nematic ordering in bilayer graphene may become
hybridized with the ordering of spin-valley symmetry-broken states. As a result, there can exist four different types
of nematic ordering. The corresponding expressions for the nematic order parameters (condensates)
can be formally obtained from those in the spin-valley symmetry-broken gapped phases by replacing
the matrix $\tau^3$ with $\tau^1$, i.e., they are given by the following anisotropic order parameters:
$\langle\Psi^{\dagger}\tau^1\Psi\rangle$,
$\langle\Psi^{\dagger}\tau^1\sigma^3\Psi\rangle$,
$\langle\Psi^{\dagger}\hat{\xi}\tau^1\Psi\rangle$, and
$\langle\Psi^{\dagger}\hat{\xi}\tau^1\sigma^3\Psi\rangle$.
These generalized nematic order parameters break the continuous spin-valley symmetry in the same
fashion as the corresponding order parameters in the QAH, QSH, QVH and LAF phases, respectively.
Also, although two of them, $\langle\Psi^{\dagger}\tau^1\sigma^3\Psi\rangle$ and
$\langle\Psi^{\dagger}\hat{\xi}\tau^1\Psi\rangle$, are invariant under time reversal,  the other two,
$\langle\Psi^{\dagger}\tau^1\Psi\rangle$ and
$\langle\Psi^{\dagger}\hat{\xi}\tau^1\sigma^3\Psi\rangle$ break this symmetry. From a physics viewpoint,
different types of nematic ordering are possible when quasiparticles with different spins and valleys
contribute unequally to the nematic condensates. In states with broken spin-valley symmetries,
this should be, of course, generally expected.

\section{Quasiparticle propagator and gap equation}
\label{secIII}

Our goal is to solve the Schwinger--Dyson (gap) equation for the quasiparticle propagator in the
spin-valley symmetry-broken gapped states as well as gapless nematic states using a
unified framework. In general, the full quasiparticle propagator in the coordinate space at fixed
valley and spin is defined by
\begin{equation}
G_{\xi s}(x-x^{\prime})=-i\langle0|T\Psi_{\xi s}(x)\Psi^{\dagger}_{\xi s}(x^{\prime})|0\rangle .
\end{equation}
In the random-phase approximation, the corresponding gap equation is readily obtained from
the Baym-Kadanoff functional in the two-loop approximation, see Eq.~(9) in the second paper in
Ref.~\onlinecite{GGM},
\begin{widetext}
\begin{eqnarray}
G^{-1}_{\xi s}(x-x^{\prime}) &=& S^{-1}_{\xi s}(x-x^{\prime})-i G_{\xi s}(x-x^{\prime}) V_{\rm eff}(x-x^{\prime})
\nonumber \\
&-& i\left[P_1(\xi)  G_{\xi s}(x-x^{\prime}) P_2(\xi) +P_2(\xi)  G_{\xi s}(x-x^{\prime}) P_1(\xi) \right]
V_{\rm IL}(x-x^{\prime})\nonumber \\
&-& \frac{i}{2} \delta^3(x-x^{\prime})
\left[ P_1(\xi)-P_2(\xi)\right] \sum_{\xi^{\prime},s^{\prime}=\pm}\mbox{tr} \left\{\left[ P_1(\xi')-P_2(\xi')\right]
G_{\xi^{\prime}s^{\prime}}(x-x^{\prime}) \right\}\tilde{V}_{\rm IL}(0) ,
\label{SD-equation}
\end{eqnarray}
\end{widetext}
where $x=(t,\mathbf{x})$ and the trace is taken over the spinor components of the quasiparticle
propagator. The dynamically screened interactions $V_{\rm eff}$ and $V_{\rm IL}$ are defined
by Eqs.~(A5) and (A6) in the second paper of Ref.~\onlinecite{GGM},
$\tilde{V}_{\rm IL}(0)=-2\pi e^2d/\kappa$ is the Fourier transform
of the bare interaction $V^{\rm bare}_{\rm IL}(x)$ taken at $\omega=\mathbf{k}=0$.
The explicit form of the inverse free propagator in momentum space reads
\begin{equation}
S^{-1}_{\xi s}(\omega,\mathbf{k})=\left(\begin{array}{cc}
\omega -\xi U_0 &
\frac{k^2}{2m}e^{-2i\varphi_k} +{\cal N}_0\\
\frac{k^2}{2m}e^{2i\varphi_k}+{\cal N}_0 &
\omega+\xi U_0
\end{array} \right) ,
\end{equation}
where $\varphi_{k}$  is a polar angle of the quasiparticle momentum with respect to the orientation of the
nematic order, $U_0=eE_{\perp}d/2$ is the top-bottom gates voltage imbalance, $E_{\perp}$ is the electric
field perpendicular to the layers, and ${\cal N}_0$ is the bare nematic order parameter due to
strain\cite{ref1} and rotational mismatch\cite{ref2} between the layers of bilayer graphene.
Its value is related, for example, to the angle of rotational mismatch $\theta$ as follows:\cite{ref2}
\begin{equation}
{\cal N}_{0} = \left(\frac{\hbar v_{F}}{2 a}\right)^2\frac{\theta^2}{\gamma_1},
\end{equation}
where $a=0.142$ nm is the intralayer distance between neighboring carbon atoms.

The Fock contribution is given by the second and third terms on the right-hand side of Eq.~(\ref{SD-equation}),
whereas, the fourth term describes the Hartree contribution. [Note that, in accordance with Gauss's
law, we omitted the Hartree contribution connected with the total electron charge of the system
proportional to $\sum_{\xi,s}\mbox{tr}[G_{\xi s}(0)]$, which is neutralized by the positive charges
of ions in the system.] Taking into account that $\langle \rho_l\rangle=-i\, \sum_{\xi,s}\mbox{tr}[P_l(\xi) \,G_{\xi s}(0)]$
defines the electron charge density in the $l$th layer, we see that the Hartree term proportional to the $\tilde{V}_{\rm IL}(0)$
interaction describes the layer charge-density imbalance.
One of the Fock terms in Eq.~(\ref{SD-equation}) is proportional to $V_{\rm IL}$. Since
$V_{\rm IL}\sim d$, it is suppressed compared to the one with $V_{\rm eff}$. Because of the
polarization effects, $V_{\rm IL}$ is also suppressed compared to $V^{\rm bare}_{\rm IL}$.\cite{QVH}
Therefore, it is justifiable to neglect the Fock term with the $V_{\rm IL}$ interaction in the analysis that follows.

Taking into account the gapped and nematic order parameters in the inverse quasiparticle
propagator at fixed valley and spin, we use the following ansatz:
\begin{widetext}
\begin{equation}
G^{-1}_{\xi s}(\omega,\mathbf{k})=\left(\begin{array}{cc}
Z^{-1}(\omega,\mathbf{k})\omega -\xi\Delta_{\xi s}(\omega,\mathbf{k}) &
\frac{k^2}{2m}A(\omega,\mathbf{k})e^{-2i\varphi_k}+w_{\xi s}(\omega,\mathbf{k})\\
\frac{k^2}{2m}A(\omega,\mathbf{k})e^{2i\varphi_k}+w_{\xi s}(\omega,\mathbf{k}) &
Z^{-1}(\omega,\mathbf{k})\omega+\xi\Delta_{\xi s}(\omega,\mathbf{k})
\end{array} \right)\,,
\label{inverse}
\end{equation}
\end{widetext}
where $k$ and $\varphi_k$ are polar coordinates in the $(k_1,k_2)$ plane. Note that
although, in general, $w_{\xi s}$ is a complex function, one can show that, without
the loss of generality,  its phase can be set to zero. The functions $Z(\omega,\mathbf{k})$
and $A(\omega,\mathbf{k})$ define quasiparticle residue at the pole and renormalization
of the kinetic term, respectively.

In Eq.~(\ref{inverse}), $\Delta_{\xi s}=U_s+\xi\Delta_s$ and $w_{\xi s} = {\cal N}_s + \xi{\cal M}_s$ are
the spin-valley symmetry-breaking and nematic parameters, which are connected to the order
parameters discussed above through the following relationship:
\begin{equation}
\langle\Psi^{\dagger}{\cal O}\Psi\rangle=-i\sum_{\xi,s}
\int\frac{d\omega d^{2}k}{(2\pi)^{3}}{\rm tr}[{\cal O} G_{\xi,s}(\omega,\mathbf{k})].
\label{condensates}
\end{equation}
Here the matrix ${\cal O}$ is $\tau^{3}$, $\tau^{3}\sigma^{3}$, $\hat{\xi}\tau^{3}$, and $\hat{\xi}\tau^{3}\sigma^{3}$
for the QAH, QSH, QVH, and LAF order parameters, respectively, and  ${\cal O}=\tau^{1}$, $\tau^{1}\sigma^{3}$,
$\hat{\xi}\tau^{1}$, and $\hat{\xi}\tau^{1} \sigma^{3}$ for nematic order parameters.
[In Eq.~(\ref{condensates}), the trace runs over the valley 
degree of freedom.] For the QSH state at $U_0=0$ and ${\cal N}_0=0$, only the spin-triplet Haldane mass
($\Delta_+=-\Delta_-$) is nonzero in $\Delta_{\xi s}$. On the other hand, for the QAH state,
only the spin-singlet Haldane mass ($\Delta_+=\Delta_-$) is nonzero. Similarly, only the
spin-antisymmetric part of $U_s$, $U_+=-U_-$, describes the LAF state, and the symmetric in spin
voltage $U_+=U_-$ occurs in the QVH state. In fact,  at $U_0=0$ these definitions remain
unchanged even in the corresponding hybrid states, in which an additional spin-singlet
nematic parameter ${\cal N}_+={\cal N}_-$ is induced by the bare nematic term ${\cal N}_0$.

When $U_0\neq 0$,
however, the most general hybrid QSH,  QAH and LAF phases get further modified. In the hybrid QSH state,
for example, there will appear a nonzero contribution of a spin-symmetric parameter $U_+=U_-$ and a
spin-antisymmetric parameter ${\cal M}_+=-{\cal M}_-$. In the hybrid QAH state, the admixture of two
spin-singlet parameters $U_+=U_-$ and ${\cal M}_+={\cal M}_-$ will appear. Finally, in the hybrid
LAF state, the new nonzero contributions will be a spin-symmetric parameter $U_+=U_-$ and a spin-antisymmetric
parameter ${\cal N}_+=-{\cal N}_-$.

Although $\Delta_{\xi s}$ and $w_{\xi s}$ can, in general, be energy and momentum dependent,
in our analysis, we use the constant approximation for them. Also, we set
$Z(\omega,\mathbf{k})=A(\omega,\mathbf{k})=1$ in $G^{-1}_{\xi s}(\omega,\mathbf{k})$
in Eq.(\ref{inverse}). Then, the propagator reads
\begin{widetext}
\begin{equation}
G_{\xi s}(\omega,\mathbf{k})=\frac{1}{\omega^2-\frac{k^4}{4m^2}-\Delta^2_{\xi s}-w^2_{\xi s}
-\frac{k^2w_{\xi s}\cos 2\varphi_k}{m}+i0}\,
\left(\begin{array}{cc} \omega+\xi\Delta_{\xi s} &
-\frac{k^2}{2m}e^{-2i\varphi_k}-w_{\xi s}\\
-\frac{k^2}{2m}e^{2i\varphi_k}-w_{\xi s} &
\omega-\xi\Delta_{\xi s}
\end{array} \right)\,.
\label{propagator}
\end{equation}
\end{widetext}
The poles of this propagator determine the dispersion relations of quasiparticles,
\begin{equation}
\omega_{\xi, s, \pm }=\pm \sqrt{\left(\frac{k^2}{2m}-w_{\xi s}\right)^2+\frac{k^2}{m}w_{\xi s}
\left(1+\cos 2\varphi_k\right)
+\Delta^2_{\xi s}} .
\end{equation}
As we see, a nonzero dynamical parameter $\Delta_{\xi s}$ results in a fully gapped dispersion
relation for the quasiparticles with fixed $\xi$ and $s$. When $\Delta_{\xi s}=0$, in contrast, the
quasiparticles with $k=\sqrt{2m |w_{\xi s}|}$  have a vanishing gap for $\varphi_k=\pm\pi/2$
and $w_{\xi s}>0$, and for  $\varphi_k=0,\pi$ when $w_{\xi s}<0$.

Multiplying the Schwinger--Dyson equation (\ref{SD-equation}) by $\tau_3$ and $\tau_1$ Pauli matrices,
respectively, taking the trace over the spinor components, and making the Wick rotation,
we obtain the following set of gap equations for $\Delta_{\xi s}$ and $w_{\xi s}$:
\begin{equation}
\Delta_{\xi s}=U_0+\int\frac{d\omega d^2k}{(2\pi)^3} \frac{\Delta_{\xi s}}{ D_{\xi s}(\omega,k) }
V_{\rm eff}(\omega,k)+ \frac{1}{2}\delta \rho \tilde{V}_{\rm IL}(0)\,,
\label{gap-equation-1}
\end{equation}
\begin{equation}
w_{\xi s}={\cal N}_{0}+\int\frac{d\omega d^2k}{(2\pi)^3} \frac{w_{\xi s}+\frac{k^2}{2m}\cos 2\varphi_k
}{ D_{\xi s}(\omega,k) } V_{\rm eff}(\omega,k),
\label{gap-equation-2}
\end{equation}
where
\begin{equation}
\delta \rho \equiv \langle \rho_2 - \rho_1\rangle =\sum_{\xi,s=\pm} \int\frac{d\omega
d^2k}{(2\pi)^3}\frac{2\Delta_{\xi s}}{ D_{\xi s}(\omega, k) }
\end{equation}
is the  charge density imbalance between the layers and
\begin{equation}
D_{\xi s}(\omega,k)=\omega^2+\Delta^2_{\xi s}+ w^{2}_{\xi s}+\frac{k^4}{4m^2}+
\frac{k^2 w_{\xi s}\cos 2\varphi_k}{m}\,.
\label{D}
\end{equation}
In the effective potential $V_{\rm eff}(\omega,k)$
in Eqs.~(\ref{gap-equation-1}) and (\ref{gap-equation-2}), we take into account the dynamical
polarization of the Coulomb interaction,\cite{QVH}
\begin{equation}
V_{\rm eff}(\omega,k)=\frac{2\pi e^2}{\kappa k+\frac{4e^2m\ln4}{\sqrt{1+[{4m\omega\ln4}/({\pi k^2})]^2}}}\,.
\end{equation}
At fixed values of $U_0$ and ${\cal N}_{0}$,
the gap equations (\ref{gap-equation-1}) and (\ref{gap-equation-2}) generally admit many
solutions. In order to determine which of them corresponds to the ground state, we need to
find the solution with the lowest energy density. The expression for the energy density is
obtained in the Appendix.

It should be mentioned that, in the model at hand, which includes only the dominant Coulomb
interaction and has no external magnetic field, the three types of solutions with spontaneous spin-valley
symmetry breaking, QAH, LAF, and QSH, appear to be degenerate in energy. Moreover, the
order parameters in these three phases are related by some transformations. This means, in
particular, that having the explicit form for one of them allows for easily reconstructing the other two.
For example, if the spin-singlet hybrid QAH solution is found to take the following form:
\begin{eqnarray}
&&\Delta_{+}= \Delta_{-} = f_1(U_0,{\cal N}_0), \\
&&U_{+} = U_{-} = f_2(U_0,{\cal N}_0), \\
&&{\cal N}_{+}=  {\cal N}_{-}= f_3(U_0,{\cal N}_0), \\
&&{\cal M}_{-} = {\cal M}_{+} = f_4(U_0,{\cal N}_0),
\end{eqnarray}
with the functions on the right-hand sides depending on $U_0$, ${\cal N}_0$ and other model
parameters, the corresponding QSH and LAF solutions can be immediately determined as well.
In particular, the QSH solution is given by
\begin{eqnarray}
&&\Delta_{+}= - \Delta_{-} = f_1(U_0,{\cal N}_0), \\
&&U_{+} = U_{-} = f_2(U_0,{\cal N}_0), \\
&&{\cal N}_{+}=  {\cal N}_{-}= f_3(U_0,{\cal N}_0), \\
&&{\cal M}_{+} =  -{\cal M}_{-} = f_4(U_0,{\cal N}_0),
\end{eqnarray}
whereas, the LAF solution is given by
\begin{eqnarray}
&&\Delta_{+}= \Delta_{-} = 0,\\
&&U_{\pm} = \frac{f_2(U_0,{\cal N}_0)\mp f_1(U_0,{\cal N}_0)}{2}, \\
&&{\cal N}_{\pm}= \frac{f_3(U_0,{\cal N}_0) \pm f_4(U_0,{\cal N}_0)}{2} , \\
&&{\cal M}_{-} = {\cal M}_{+} = 0\,.
\end{eqnarray}

\section{Results}
\label{secIV}
In this section, we present our main results. We start from the analysis of the gap equations
in the case of vanishing top-bottom gates voltage imbalance, $U_0=0$. In particular, we reveal the
emergence of the hybrid state at nonzero bare nematic parameter ${\cal N}_{0}$ and describe how
it gradually turns into a pure nematic state with increasing ${\cal N}_{0}$. We then discuss the
complete phase diagram in the $U_0$--${\cal N}_{0}$ plane.

\subsection{Nematic and spin-valley symmetry-broken states at $U_0=0$}

In the hybrid QSH, QAH, and LAF states (with an addition of nematic ordering),
we find that in general the nematic parameters depend on spin and/or valley indices,
i.e., ${\cal N}_{+}\neq {\cal N}_{-}$ and ${\cal M}_{s} \neq 0$. However, in the case
of the vanishing top-bottom gates voltage imbalance, $U_0=0$, the structure of $w_{\xi s}$
simplifies: ${\cal N}_{+} = {\cal N}_{-}$ and ${\cal M}_{s} = 0$. The latter follows
from the structure of Eqs.~(\ref{gap-equation-1}), (\ref{gap-equation-2}), and (\ref{D}).
In order to clarify the role of a bare nematic parameter ${\cal N}_{0}$ in competition between
the nematic and the gapped states, we first study this case.
Then, the numerical analysis of gap equations (\ref{gap-equation-1}) and (\ref{gap-equation-2})
shows that solutions of the two types are possible. One of them is the nematic solution with only
$w_{\xi s}\ne0$, and the other is a hybrid solution with both $\Delta_{\xi s}$ and $w_{\xi s}$ nonzero.
Note that a gapped solution with only $\Delta_{\xi s} \ne 0$ is impossible when a nonzero bare nematic
parameter ${\cal N}_{0}$ is present, therefore, gap and nematic parameters {\it coexist}
in the hybrid solution.

At vanishing $U_0$ and ${\cal N}_0$, the QVH solution has a higher energy than the QAH, LAF, and QSH
solutions. The additional energy cost is due to the Hartree term $\frac{1}{2} \delta \rho \tilde{V}_{\rm IL}$
in the gap equation (\ref{gap-equation-1}). This term is negative because it reflects the energy price 
associated with the electric field between the layers of graphene in the QVH phase. Thus, there are 
two main solutions of the gap equations (\ref{gap-equation-1}) and (\ref{gap-equation-2}), the pure 
nematic and one of the hybrid solutions, QAH, LAF, or QSH. The latter are degenerate
in energy. Instead of discussing all the degenerate solutions, in the rest of the paper, we will
concentrate only on the hybrid QSH one.

\begin{figure}[t]
\includegraphics[scale=0.44]{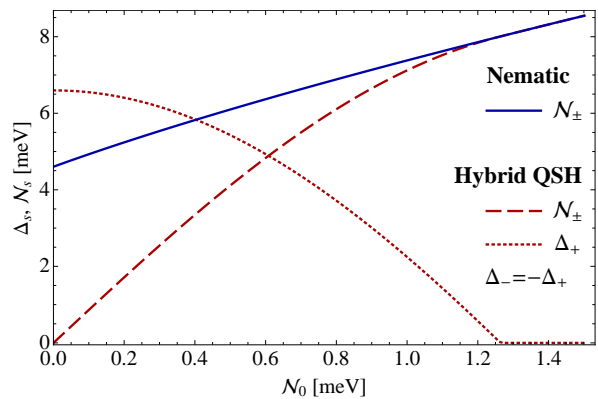}
\caption{(Color online)
The order parameters $\Delta_s$ and ${\cal N}_{s}$ as functions of the bare nematic term
${\cal N}_{0}$ for the nematic solution (solid line) and for the hybrid QSH solution (dashed
and dotted lines) at $U_0=0$.}
\label{fig1}
\end{figure}

In the pure nematic solution, the nematic order parameter ${\cal N}_{s}$ does not depend on the
valley or the spin. Its dependence on the bare parameter ${\cal N}_{0}$ is shown in Fig.~\ref{fig1}.
At small ${\cal N}_{0}$, it can be approximated by a linear dependence ${\cal N}_{\pm}={\cal N}_{\rm off}+b{\cal N}_{0}$
with the offset ${\cal N}_{\rm off}=4.61\, \mbox{meV}$ and the slope $b=3.18$. For the hybrid QSH solution,
the gaps $\Delta_{s}$ and ${\cal N}_{s}$ are also plotted in Fig.~\ref{fig1}.
We see that the bare nematic parameter inhibits the gap.

%%%%%%%%%%%%%%%%%%%%%%%%%%%%%%%%%%%%%%%
%%%%%%%%%%%%%%%%%%%%%%%%%%%%%%%%%%%%%%%

\begin{figure}[t]
\includegraphics[scale=0.48]{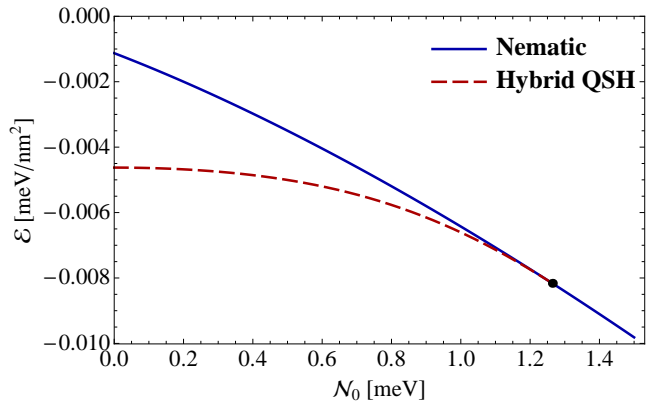}
\caption{(Color online)
The energy density as a function of the bare nematic term ${\cal N}_{0}$ for the
nematic solution (solid line) and the hybrid QSH solution (dashed line) at $U_0=0$.}
\label{figEnergy}
\end{figure}

Comparing the energy densities of the nematic and hybrid QSH solutions, we find that
the hybrid solution always has a lower free energy whenever it exists, see Fig.~\ref{figEnergy}.
From Fig.~\ref{fig1}, we see that the dynamical nematic order parameter ${\cal N}_s$
for the hybrid QSH solution vanishes as ${\cal N}_{0} \to 0$. Therefore, the ground
state of unbiased bilayer graphene at ${\cal N}_{0}=U_{0}=0$ is a spin-valley
symmetry-broken gapped state. Whereas, the dynamical nematic order parameter ${\cal N}_s$
grows, the gap $\Delta_{s}$ gradually decreases with increasing the bare nematic parameter
${\cal N}_{0}$. Eventually, the hybrid QSH solution smoothly turns into the pure nematic solution
at the critical point ${\cal N}^{\rm cr}_{0} \approx 1.26\, \mbox{meV}$ where the gap $\Delta_{s}$
turns to zero. Since the QSH state breaks the spin $SU(2)$ symmetry, whereas, the nematic state
does not, we conclude that the corresponding transition is
a second-order phase transition (and not a smooth crossover).

%%%%%%%%%%%%%%%%%%%%%%%%%%%%%%%%%%%%%%%
%%%%%%%%%%%%%%%%%%%%%%%%%%%%%%%%%%%%%%%

\subsection{Phase diagram}

\begin{figure}[t]
\includegraphics[width=0.44\textwidth]{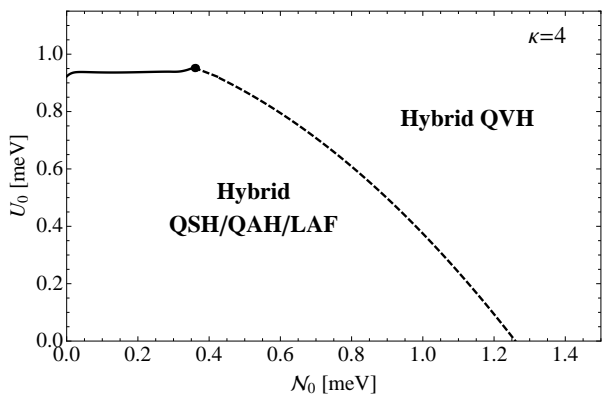}
\caption{The phase diagram in the ${\cal N}_{0}$-$U_{0}$ plane.}
\label{phase-diagram}
\end{figure}

Using a coupled set of gap equations for the nematic and several types of spin-valley
symmetry-breaking order parameters in bilayer graphene, we can now obtain a phase
diagram in the plane of two parameters: a strain-induced bare nematic term ${\cal N}_{0}$
and a top-bottom gates voltage imbalance $U_{0}$. This is performed by sweeping through the
corresponding two-dimensional space of the bare parameters, comparing the energies of all
solutions, and determining the ground state at each point. The results are summarized in
Fig.~\ref{phase-diagram}.

%%%%%%%%%%%%%%%%%%%%%%%%%%%%%%%%%%%%%%%
%%%%%%%%%%%%%%%%%%%%%%%%%%%%%%%%%%%%%%%

To understand the overall topology of the phase diagram, it is useful to first look at the
competition of the non-nematic phases at ${\cal N}_0=0$.\cite{GGM,Levit,Toke} In this case,
increasing the voltage imbalance $U_0$ has a tendency to suppress the QSH state gap
$\Delta_s$ and to induce an increasing admixture of the $U_s$ gap in this state.
At a sufficiently large $U_0$, a pure QVH state with a spin singlet $U_s$ gap takes
over as the ground state.

\begin{figure}[t]
\includegraphics[width=0.44\textwidth]{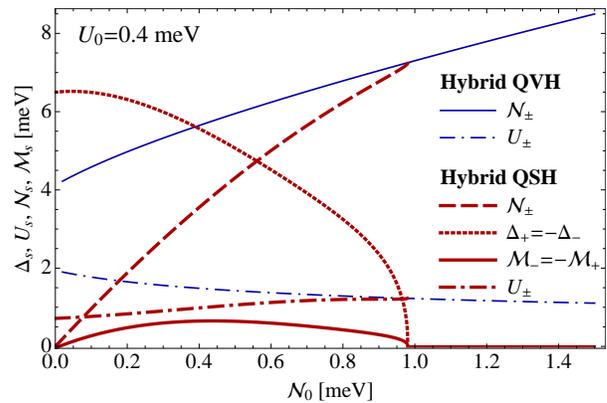}
\caption{(Color online)
The order parameters as functions of the bare nematic term ${\cal N}_{0}$ for the
hybrid QSH solution (thick red lines) and hybrid QVH solution (thin blue lines) at $U_0=0.4\,\mbox{meV}$.}
\label{results-fixed-U0}
\end{figure}

In the subcritical region (i.e., at small values of ${\cal N}_{0}$ and $U_{0}$) of
the phase diagram at a fixed $U_0\neq 0$, increasing the bare nematic term ${\cal N}_{0}$
induces a substantial dynamical nematic order parameter $w_{\xi s}$, see Fig.~\ref{results-fixed-U0}
for $U_0=0.4\,\mbox{meV}$.
Consequently, in this case, the hybrid QSH state has spin-symmetric order parameters $U_s$ and
${\cal N}_s$, and spin asymmetric order parameters $\Delta_s$ and ${\cal M}_s$.
In the supercritical region, the hybrid QVH state with spin singlet $U_s$ and ${\cal N}_s$ parameters
(see Fig.~\ref{results-fixed-U0}) has the lowest energy. As expected, the hybrid QVH state
becomes almost a pure nematic state at small $U_0$
and large ${\cal N}_{0}$ and almost a pure QVH state at large $U_0$ and small ${\cal N}_{0}$.

%%%%%%%%%%%%%%%%%%%%%%%%%%%%%%%%%%%%%%%
%%%%%%%%%%%%%%%%%%%%%%%%%%%%%%%%%%%%%%%

The critical line in Fig.~\ref{phase-diagram} separates the hybrid QSH and QVH phases.
Its top part corresponds to a first-order phase transition (the gaps of the hybrid QSH and QVH states
are discontinuous across the critical line), which ends in a critical end point at
${\cal N}^{\star}_{0}\approx 0.36\, \mbox{meV}$
and $U^{\star}_{0}\approx 0.95\, \mbox{meV}$. The rest of the critical line corresponds to a
second-order phase transition.\\

\section{Summary}
\label{Summary}

The current paper shows that bilayer graphene has a rich phase diagram in the plane of the
bare nematic term ${\cal N}_{0}$ and the voltage imbalance between the layers $U_0$.\cite{footnote2}
It is argued that the ground state should be one of several hybrid states with spontaneous
spin-valley symmetry breaking.

We find, in particular, that one of the hybrid states, QAH, LAF, or QSH state is the ground state 
in the subcritical region
(i.e., small values of ${\cal N}_{0}$ and $U_0$), and that a hybrid QVH state is the ground state
in the supercritical region (i.e., large values of ${\cal N}_{0}$ and/or $U_0$). The transition
between the two phases is either a first-order phase transition (at small ${\cal N}_{0}$ and
large $U_0$), or a  second-order phase transition. The two parts of the critical line meet at
a critical end point, which is an interesting prediction in its own right.

Note that, in the model with the Coulomb interaction used in this paper, three of the proposed
hybrid states (i.e., QAH, LAF, and QSH) appear to be degenerate in energy. This degeneracy
is expected to be lifted after the inclusion of additional symmetry-breaking contact interactions.
Indeed, as follows from the renormalization-group studies in the normal phase,\cite{Vafek,
Lemonik,Aleiner,1205.5532} there are several
types of such interactions that may become relevant for the low-energy dynamics. The
corresponding generalization of the phase diagram is of great interest and will be addressed
in the future.\cite{Work-in-prog}

Based on the general arguments reviewed in Ref.~\onlinecite{Hasan}, one should expect
that, although gapless edge states are present in (hybrid) QAH and QSH states, there are
no such states in (hybrid) QVH and LAF ones. The role of the edge states in
present dynamics will be considered elsewhere.

It is interesting to note that certain two-dimensional fermionic models with a quadratic
band-crossing point similar to bilayer graphene, but with a $d$-wave symmetry, can
also have a hybrid phase with coexisting spin-valley symmetry-breaking and nematic
order parameters.\cite{Fradkin}

The experimental data in Ref.~\onlinecite{Mayorov} were fitted with the nematic parameter on the
order of $6~\mbox{meV}$, and it was argued that the required strain to explain the observed effects
should be too large. According to our results, interactions strongly enhance the bare nematic
parameter (e.g., ${\cal N}/{\cal N}_0\simeq 6.3$ for ${\cal N}_0\simeq {\cal N}^{\rm cr}_0$
at $U_0=0$). Therefore, even a rather small strain can lead to the nematic state. This
suggests that different bilayer samples in experiments may have very different physical
properties, depending on the specific conditions under which they were created.

\begin{acknowledgements}
The work of E.V.G. and V.P.G. was supported partially  by the Scientific Cooperation
Between Eastern Europe and Switzerland (SCOPES) program under Grant
No.~IZ73Z0\_128026 of the Swiss National Science Foundation (NSF),
the European FP7 program, Grant No. SIMTECH 246937, and the joint Ukrainian-Russian
SFFR-RFBR Grant No.~F40.2/108. V.P.G.  acknowledges a collaborative grant from the
Swedish Institute. The work of V.A.M. was supported by the Natural Sciences and
Engineering Research Council of Canada. The work of I.A.S. was supported, in part, by
the U.S. National Science Foundation under Grant No.~PHY-0969844.
\end{acknowledgements}

\appendix
\section{Energy density}
\label{energy-density-derivation}

According to Eq.~(2.9) in Ref.~\onlinecite{KB}, the energy density of a three-dimensional electron
gas can be calculated through an integral of the electron Green's function. It is not difficult
to check that, with obvious modifications, this formula is valid also for electron quasiparticles in bilayer graphene.
Alternatively, the energy density can be obtained by evaluating the Baym-Kadanoff functional
on the solutions of the gap equations. Then, we have
\begin{widetext}
\begin{equation}
{\cal E}=-\frac{i}{2}\sum_{\xi,s=\pm}\int\frac{d\omega}{2\pi} \int
\frac{d^2k}{(2\pi)^2}\,\,\mbox{tr}\left[\,(\omega+\xi U_0\tau_3+{\cal N}_0\tau_1+H_0)\,
G_{\xi s}(\omega,\mathbf{k})\,\right],
\label{energy-density}
\end{equation}
where $H_0$ is the free Hamiltonian of bilayer graphene. Using the expression for the propagator
in Eq.~(\ref{propagator}), we derive an explicit expression for the  energy density,
\begin{equation}
{\cal E}=-i\sum_{\xi,s=\pm}\int\frac{d\omega}{2\pi}\int\frac{d^2k}{(2\pi)^2}\,
\frac{\omega^2+U_0\Delta_{\xi s}+\frac{k^4}{4m^2}+\frac{k^2({\cal N}_0+w_{\xi s})
\cos 2\varphi_k}{2m}+{\cal N}_0w_{\xi s}}{\omega^2-\Delta^2_{\xi s}-w^2_{\xi s}-\frac{k^4}{4m^2}
-\frac{k^2w_{\xi s}\cos 2\varphi_k}{m}+i0}\,,
\label{energy-density1}
\end{equation}
which can be used for any state with dynamically generated gaps and nematic parameters.
Note that the integral over $\omega$ diverges in this expression for ${\cal E}$. Therefore, we should
subtract the ``vacuum" energy of the trivial solution with $\Delta_{\xi s}=w_{\xi s}=0$. Then,
we obtain
\begin{equation}
{\cal E}=-i\sum_{\xi,s=\pm}\int\frac{d\omega}{2\pi}\int\frac{d^2k}{(2\pi)^2}\,\left[\,
\frac{\omega^2+U_0\Delta_{\xi s}+\frac{k^4}{4m^2}+\frac{k^2({\cal N}_0+w_{\xi s})\cos 2\varphi_k}{2m}
+{\cal N}_0w_{\xi s}}
{\omega^2-\Delta^2_{\xi s}-w^2_{\xi s}-\frac{k^4}{4m^2}-\frac{k^2w_{\xi s}\cos 2\varphi_k}{m}+i0}-
\frac{\omega^2+\frac{k^4}{4m^2}}
{\omega^2-\frac{k^4}{4m^2}+i0}
\,\right]\,.
\label{energy-density2}
\end{equation}
Performing the Wick rotation and integrating over $\omega$, we arrive at the expression,
\begin{equation}
{\cal E} = -\frac{1}{2}\sum_{\xi, s=\pm}\int\frac{d^2k}{(2\pi)^2} \left[
\frac{\Delta^2_{\xi s}+U_0 \Delta_{\xi s}+w^2_{\xi s}+ {\cal N}_{0} w_{\xi s} }
{\sqrt{\Delta^2_{\xi s}+w^2_{\xi s}+\frac{k^4}
{4m^2}+\frac{k^2w_{\xi s}\cos 2\varphi_k}{m}}}
+ \frac{\frac{k^4}{2m^2}
+\frac{k^2(3w_{\xi s}+{\cal N}_{0})\cos 2\varphi_k}{2m}}
{\sqrt{\Delta^2_{\xi s}+w^2_{\xi s}+\frac{k^4}
{4m^2}+\frac{k^2w_{\xi s}\cos 2\varphi_k}{m}}}
-\frac{k^2}{m}
\right].
\label{energy-density-general}
\end{equation}
\end{widetext}
The integral over the momentum in this expression is logarithmically divergent
at large $k$. In the calculations, therefore, we use a finite cutoff at $k_{\rm max}=\sqrt{2m \Lambda}$,
which is determined by the range of validity of the effective Hamiltonian (\ref{free-Hamiltonian})
(recall that $\Lambda=\gamma_1/4$).


\begin{thebibliography}{99}
\bibitem{footnote2}
It is instructive to compare the present paper with that of Ref.~\onlinecite{Aleiner}. They are
quite different: Whereas, bilayer graphene in the plane of control parameters ${\cal N}_{0}$
and $U_0$ is studied here, the phase diagram in the plane of coupling constants is
investigated in Ref.~\onlinecite{Aleiner}.
\end{thebibliography}
\end{document}